# Horizon-T Experiment Calibrations - Cables

D. Beznosko [1b], A. Iakovlev [b], Z. Makhataeva [b], M.I. Vildanova [a], V.V. Zhukov [a]

*Abstract*

An innovative detector system called Horizon-T is constructed to study Extensive Air Showers (EAS) in the energy range above $10^{16}$ eV coming from a wide range of zenith angles ($0^o$ - $85^o$). The system is located at Tien Shan high-altitude Science Station of Lebedev Physical Institute of the Russian Academy of Sciences at approximately 3340 meters above the sea level.

The detector consists of eight charged particle detection points separated by the distance up to one kilometer as well as optical detector to view the Vavilov-Cherenkov light from the EAS. Each detector connects to the Data Acquisition system via cables. The calibration of the time delay for each cable and the signal attenuation is provided in this article.

## 1. Detector System Description

The high-altitude Science Station Tien Shan is a branch of the Lebedev Physical Institute of the Russian Academy of Science. It is located 32 km from Almaty at the altitude of 3340 meters above the sea level. “Horizon-T” detector system [1] [2] is used to study EAS with parent particle of energies higher than $10^{16}$ eV coming from a range of zenith angels ($0^o$-$85^o$). “Horizon-T” is constructed to study space-time distribution of the charged particles in EAS disk and Vavilov-Cherenkov radiation from it. The novel method of using time information from pulse shape in each detector allows for the analysis of EAS with core falling outside of the detector system bounds.

The EAS disk passes the level of observation in a time range from ~10ns to ~100ns, depending on the distance from its center. Time of passage of the charged particles disk through the setup is registered at eight detection points. The relative coordinates of every point are presented in the Table 1 with the aerial view of the detector system in Figure 1. The system center is indicated by a geodesic benchmark installed at the detection point 1 at the height of 3346.05 meters above the sea level and with geographical coordinates of 43°02′49.1532" N and 76°56′43.548" E. This benchmark is the origin for the XYZ coordinate system for Horizon-T. The X-axis is directed to the north, Y-axis to the west and Z-axis is directed vertically up. The geometric factor of the detector system is 1 $km^2$/ster. The detector of Vavilov-Cherenkov radiation is located next to station 1. The signal-carrying cables of all three Cherenkov detectors are of the same length. Plans exist to build a cable-less version of the detector system [3].

[1] dmitriy.beznosko@nu.edu.kz (also dima@dozory.us)
[a] P. N. Lebedev Physical Institute of the Russian Academy of Sciences, Moscow, Russia
[b] Physics Department, Nazarbayev University, Astana, Kazakhstan

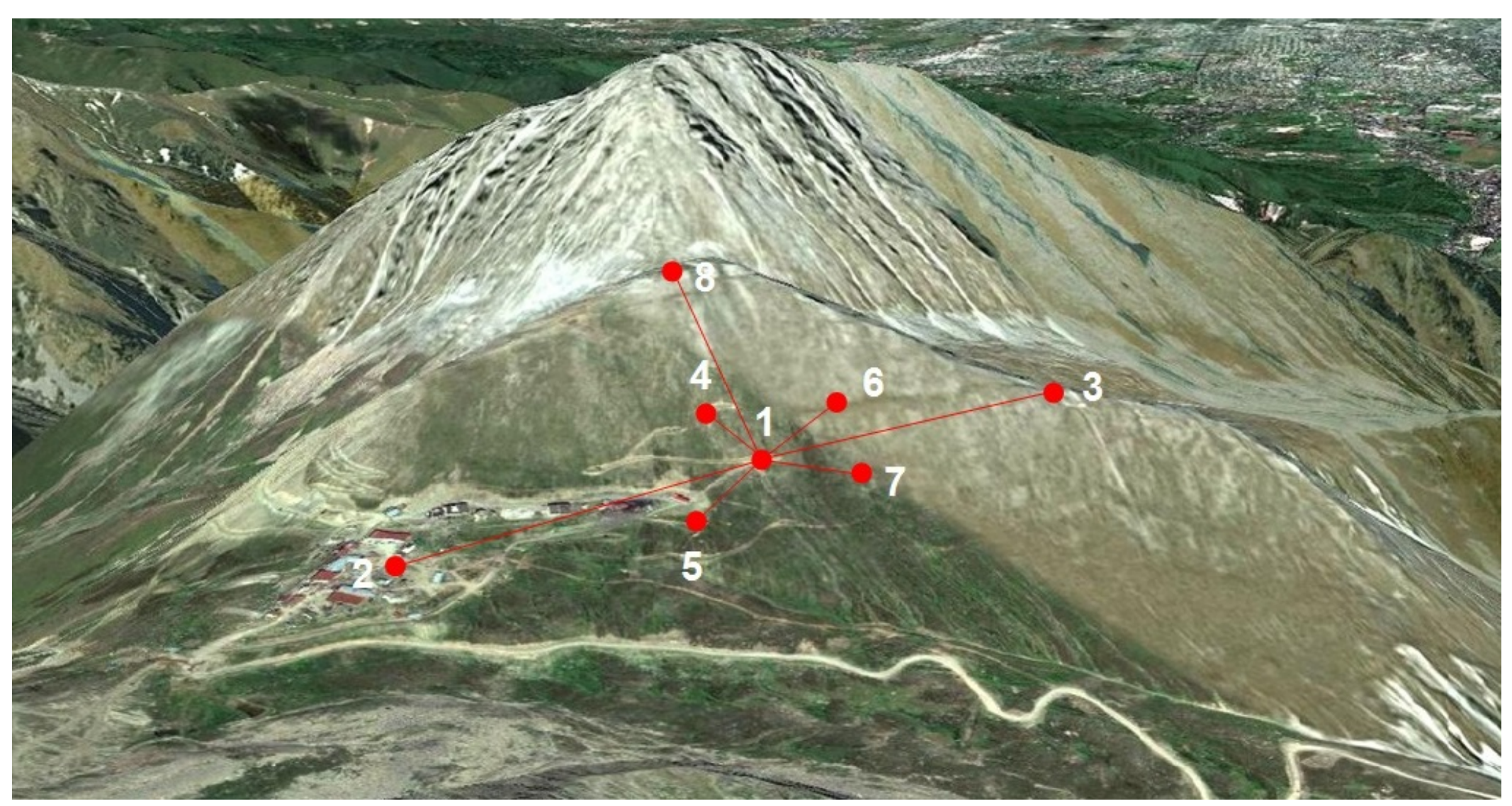

**Figure 1: Detection points aerial view.**

**Table 1: Coordinates of all detection points.**

| Station # | X, m | Y, m | Z, m | R, m |
|---|---|---|---|---|
| 1 | 0 | 0 | 0 | 0 |
| 2 | –445.9 | –85.6 | 2.8 | 454.1 |
| 3 | 384.9 | 79.5 | 36.1 | 394.7 |
| 4 | –55.0 | –94.0 | 31.1 | 113.3 |
| 5 | –142.4 | 36.9 | –12.6 | 147.6 |
| 6 | 151.2 | –17.9 | 31.3 | 155.4 |
| 7 | 88.6 | 178.4 | –39.0 | 203.0 |
| 8 | 221.3 | 262.0 | 160.7 | 378.7 |

## 1.1 Scintillator Detectors

Three scintillator detectors (SD), oriented perpendicular to each other in the x, y and z planes, are located at each detection point except point 8 that has only the z-plane one. The z-plane is parallel to the sky. This arrangement is needed for the angular isotropy in the registration of charged particles. Theoretically, better isotropy may be achieved by an upgrade to liquid scintillators [4] [5] with a symmetric active volume but it is not being yet considered at this time.

Each SD uses polystyrene-based square-shaped cast scintillator [6] with 1 m$^2$ area and 5 cm width. Light is registered either by photoelectrical multiplier PMT-49 (FEU49) from MELZ [7] that is a 15cm diameter spherical-shaped cathode PMT with the spectral response from 360nm to 600nm, or the 2-inch Hamamatsu [8] R7723 PMT. Two signals come from each detection point to the data acquisition system (DAQ) that is located at station 1: one from the horizontally installed (in z-plane) SD, another is a sum of the two vertical SDs (in x-y planes). All PMT signals are carried without any amplifiers over the coaxial cables RK 75-7-316F-C SUPER produced by SpetsKabel [9], and are also connected to the trigger board that forms the master trigger signal. From the 8 detection points, there are 15 signal-carrying cables plus three additional ones used for testing purposes at the detection point 1.

### 1.2 Cherenkov Detector

The Vavilov-Cherenkov Detector (VCD) is located next to detection point 1 as close to DAQ system as possible. The VCD consists of three parabolic mirrors 150cm diameter and 65cm focal length each mounted on the rotating support allowing registration of radiation in zenith angle range of $0^{o}$-$80^{o}$ and azimuthal angle range of $0^{o}$ -$360^{o}$. There is a PMT-49B (FEU49B) and a Hamamatsu H6527 PMT located in the focal point of two lower mirrors. The photocathode diameter for both PMTs is 15cm, thus allowing for the angle of view for each mirror is ~$13^{o}$. A considered upgrade for this system may include the Geiger-mode avalanche photodetector [10] [11] arrays as they are not damaged by high light intensity such as moonlight. These are fast detectors as well and have been already used on a large scale in [12]. The three VCD channels are connected to DAQ using separate cable each.

## 2. Calibration of Cables

All PMTs are connected to the DAQ system via coaxial cables of length ranges from 20m to ~500m. All these cables require calibration of the time delay and signal attenuation. In addition, the PMT pulse experiences widening as it traverses the cable. Note that attenuation of the signal is not used separately as the single particle response calibration of the SD is done via the actual cables thus it includes the attenuation. But the cable testing still provides the attenuation as the ratio of the reflected pulse area to the original one. The square root of the ratio should be taken to obtain the attenuation over the single length of the cable since with the use of reflection method the pulse travels the double length of the cable.

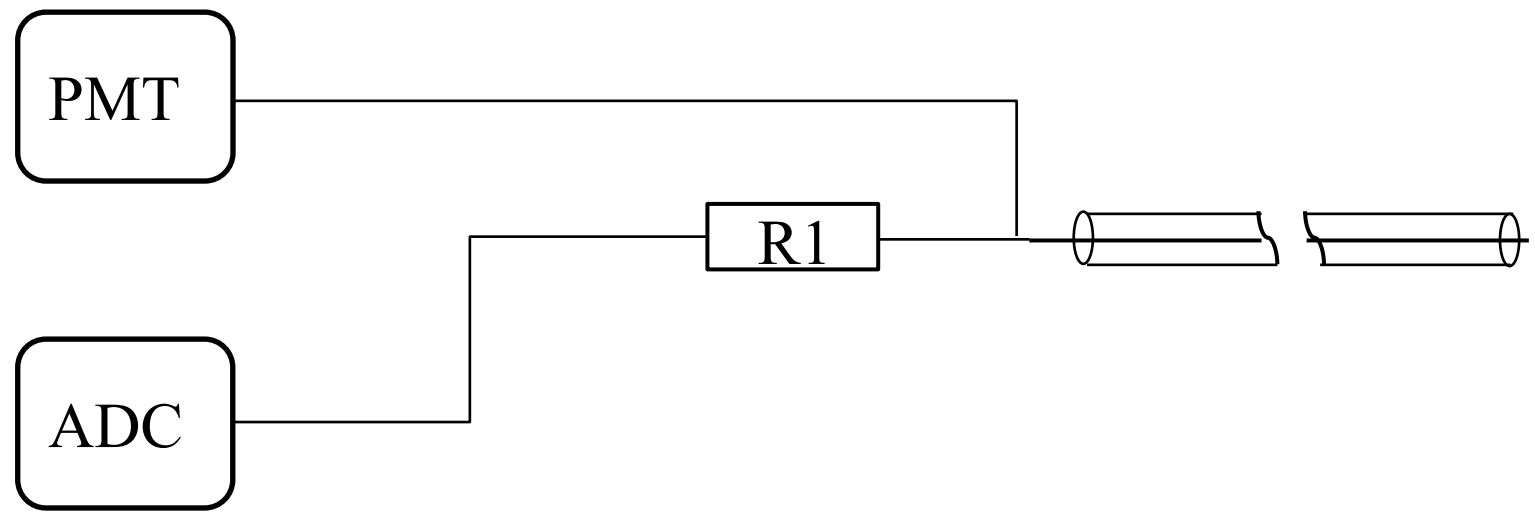

**Figure 2: cable calibration setup schematic: ADC is used with PMT as signal source.**

The common setup used is shown in Figure 2. The cable is disconnected at one end to obtain pulse reflection from it. R1 is the 25Ω resistor to achieve the impedance matching between the 50Ω for the ADC, and the cable that has impedance of 75Ω.

A test signal was provided from a stand-alone PMT49 with a 15cm x 15cm x 1cm plastic scintillator piece, and is recorded by the 14bit CAEN [13] DT5730 flash ADC. The PMT was biased at -1500V and cosmic ray events were selected by 40mV threshold.

## 2.1 Signal Analysis

The waveform recorded by the ADC consist of 5110 data points digitized every 2ns each, for the total of 10.22 μs. Since it is more convenient to work with positive polarity signals, the waveforms are inverted by being subtracted from $2^{14}$ as 0-level is set there during the measurements. The average of the first hundred points is used to set the baseline value that is also subtracted from the waveform to zero the baseline. Figure 3 shows the (zoomed) baseline-subtracted inverted waveform showing the original pulse and the reflection.

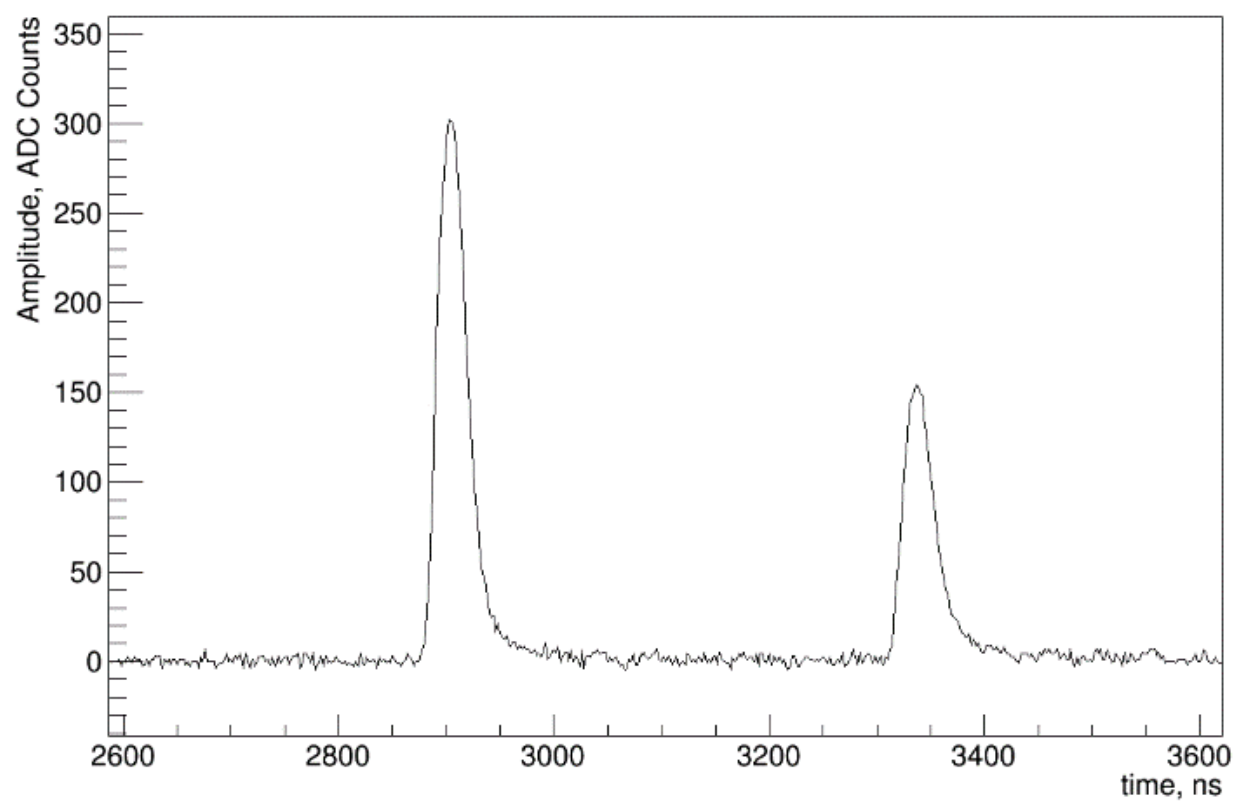

**Figure 3: Calibration signal waveform, inverted, baseline subtracted and zoomed.**

As the pulse widens, it is not trivial to define the time delay between the original and the reflected pulses. Normally, this definition is analysis-specific. For analysis purposes, the pulse front and total duration are defined as following: the *pulse front* is between 0.1 and 0.5 of the total area under the pulse (e.g. of the pulse cumulative distribution, CDF) and the *total duration* is between the level 0.1 and level 0.9 of the pulse cumulative distribution. Therefore, the time between the original and the reflected pulse is calculated at the same area values as well (e.g. 0.1, 0.5 and 0.9 of thepulse area). To illustrate this using Figure 4, for time difference at 0.5 area, the difference in x-position is calculated between the points, corresponding to y-values of 0.5 for each pulse.

Additionally, ratios of total of the reflected pulse and the area of the original pulse are also calculated to monitor the charge loss in cable.

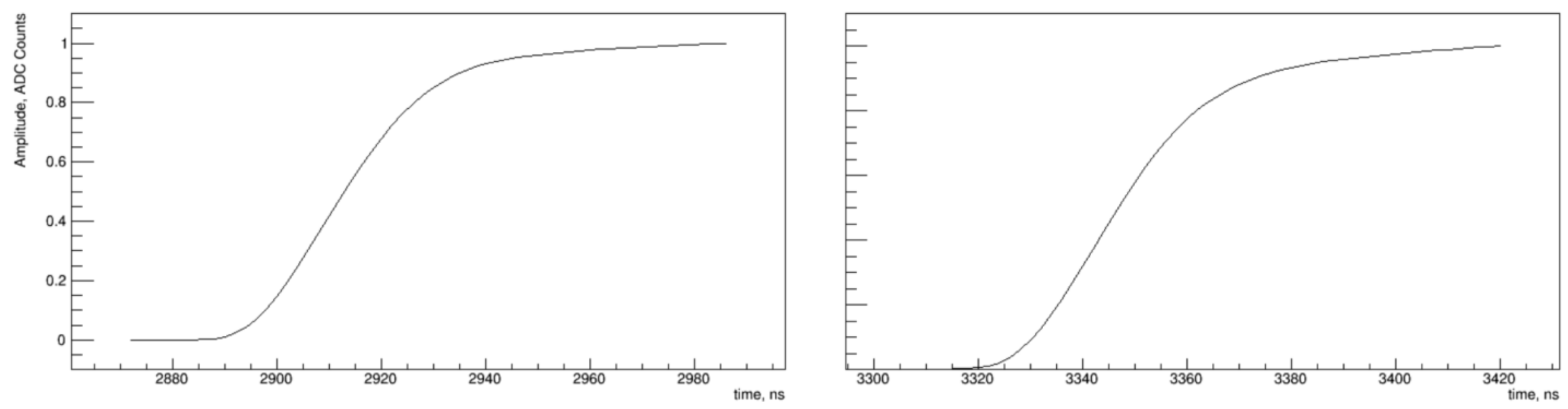

**Figure 4: Normalized to unity integrals for original (left) and reflected (right) pulses.**

First, the ADC data is processed. The integral is computed using the trapezoidal rule from the numpy package for Python programming language [14]. The integral of each pulse with the area normalized to unity as illustrated in Figure 4. For all calibration pulses (~100 per each cable), the time between the mentioned above area fractions is computed. In addition, the values of areas themselves are also stored and compared to see the signal losses in cable.

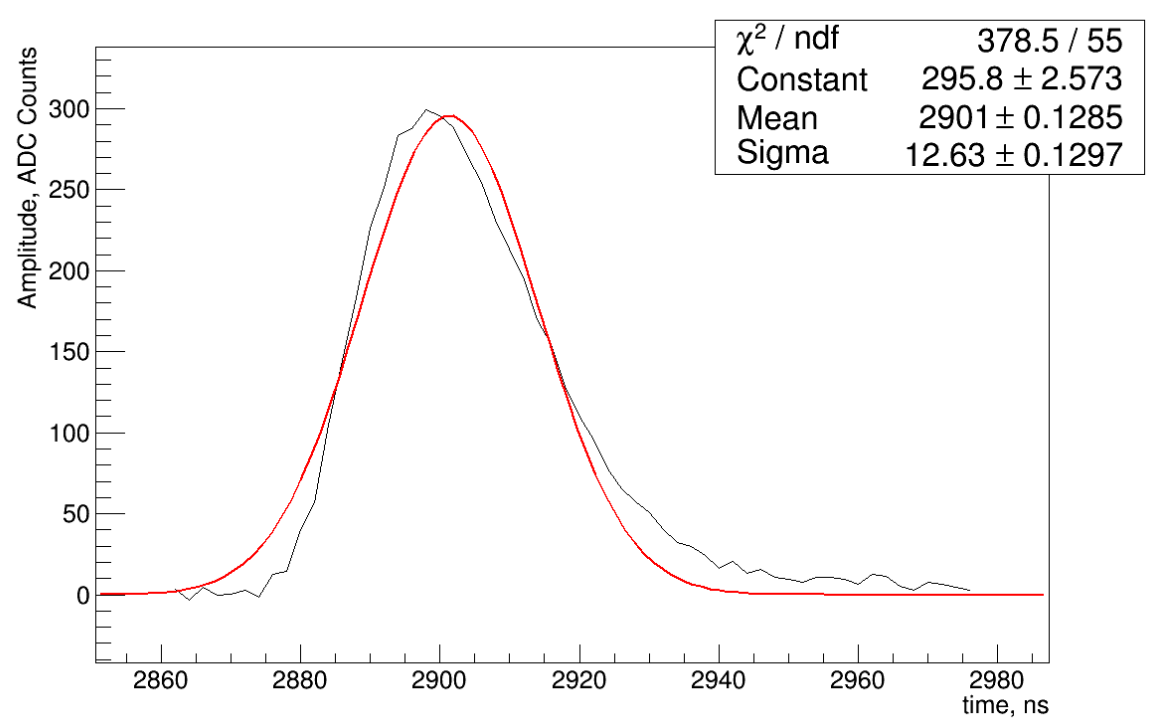

**Figure 5: Example of Gaussian fit to the original peak.**

Since defining pulse integration range is not straightforward in the case where not all pulses are identical, one needs to define integration range consistently for all pulses. This is achieved by the following schema: first, the peak of the pulse is found as a simple maximum. Then, a Gaussian is fitted to the peak. This fit (shown in Figure 5) is not matching the pulse well, but it is very stable and gives the approximate width of the pulse in a consistent method for all pulses encountered. Then integration range is defined as $-4\sigma$ and $+8\sigma$ from that the mean. Range is not symmetric to cover for the pulse intrinsic asymmetry. After the fit, the integral for each peak is numerically found. The next chapter will include the results from this calibration. Fit is done using ROOT package imported into Python (e.g. PyROOT) [15].

## 2.2 Data Analysis Results

The stations, besides having numbers, also have their 'internal' names as well as color codes for the cables. The calibration data is given using the names for the convenience. Table 2 gives the connection between the station numbers and the internal names.

**Table 2: Names of detection stations**

| *Station name* | Center | Yastrebov | Stone Flower | Left | Kurashkin | Right | Bottom | Upper | Cher |
|---|---|---|---|---|---|---|---|---|---|
| *Station number* | 1 | 2 | 3 | 4 | 5 | 6 | 7 | 8 | VCD |

The time calibration values and the ratios of the areas obtained from the ADC data directly are given in the Table 3. The pulse widening is noticeable from the simple comparison of the time differences at 10, 50 and 90 percent of the pulse area. Specifically, this is very pronounced for the farthest stations, such as 2, 3 and 8.

**Table 3: Time differences and area ratios from ADC data**

| Station and cable | Time diff. at 10% (ns) | Time diff. at 50% (ns) | Time diff. at 90% (ns) | Area ratio |
|---|---|---|---|---|
| Bottom New | 764.98 ± 0.10 | 767.31 ± 0.15 | 783.69 ± 0.86 | 0.479 ± 0.004 |
| Bottom Old | 1016.14 ± 0.11 | 1020.76 ± 0.17 | 1040.95 ± 0.65 | 0.637 ± 0.005 |
| Center Blue | 73.14 ± 0.08 | 73.57 ± 0.06 | 76.24 ± 0.31 | 0.598 ± 0.003 |
| Center Blue New | 102.14 ± 0.08 | 103.28 ± 0.08 | 108.39 ± 0.45 | 0.719 ± 0.005 |
| Center Green | 73.14 ± 0.08 | 74.176 ± 0.06 | 76.56 ± 0.36 | 0.598 ± 0.003 |
| Center Red | 73.16 ± 0.06 | 73.52 ± 0.05 | 75.71 ± 0.34 | 0.597 ± 0.004 |
| Center Red New | 102.35 ± 0.08 | 103.49 ± 0.08 | 108.83 ± 0.50 | 0.72 ± 0.005 |
| Center White | 73.56 ± 0.07 | 73.97 ± 0.06 | 76.51 ± 0.33 | 0.596 ± 0.003 |
| Center Yellow | 73.09 ± 0.06 | 73.43 ± 0.05 | 75.98 ± 0.30 | 0.596 ± 0.003 |
| Cher Green Red | 88.12 ± 0.09 | 88.91 ± 0.06 | 92.63 ± 0.43 | 0.583 ± 0.004 |
| Cher White Blue | 82.18 ± 0.12 | 83.25 ± 0.07 | 87.47 ± 0.41 | 0.583 ± 0.004 |
| Cher Yellow | 81.42 ± 0.11 | 82.46 ± 0.08 | 86.44 ± 0.62 | 0.582 ± 0.003 |
| Kurashkin New | 710.03 ± 0.10 | 712.56 ± 0.12 | 723.45 ± 0.61 | 0.506 ± 0.004 |
| Kurashkin Old | 893.38 ± 0.08 | 895.50 ± 0.11 | 905.31 ± 0.60 | 0.521 ± 0.003 |
| Left New | 552.59 ± 0.22 | 555.07 ± 0.21 | 565.50 ± 1.40 | 0.525 ± 0.008 |
| Left Old | 1061.06 ± 0.29 | 1064.19 ± 0.31 | 1077.26 ± 1.60 | 0.511 ± 0.008 |
| Right New | 561.96 ± 0.08 | 564.00 ± 0.11 | 574.15 ± 0.60 | 0.526 ± 0.003 |
| Right Old | 717.57 ± 0.11 | 720.83 ± 0.16 | 735.60 ± 0.96 | 0.475 ± 0.005 |
| Right Old2 | 778.86 ± 0.15 | 780.61 ± 0.16 | 791.57 ± 1.14 | 0.337 ± 0.007 |
| Stone Flower New | 1634.06 ± 0.27 | 1642.09 ± 0.37 | 1675.85 ± 2.40 | 0.429 ± 0.01 |
| Stone Flower Old | 2332.60 ± 0.26 | 2339.77 ± 0.42 | 2368.64 ± 1.70 | 0.461 ± 0.008 |
| Yastrebov New | 2079.86 ± 0.48 | 2090.90 ± 0.90 | 2134.78 ± 5.69 | 0.409 ± 0.029 |
| Yastrebov Old | 2548.12 ± 0.36 | 2556.63 ± 0.49 | 2590.20 ± 2.60 | 0.451 ± 0.011 |
| Upper New using function generator | 2049.13 ± 0.05 | 2057.35 ± 0.05 | 2094.28 ± 0.21 | 0.337 ± 0.0008 |
| Upper New | 2050.39 ± 1.16 | 2061.21 ± 1.50 | 2108.40 ± 5.70 | 0.369 ± 0.031 |

## 3. Conclusion

The calibration of all cables of Horizon-T has been completed. The time delay of the pulse at three different area fraction points is done for each pulse, thus showing the pulse widening as it travels along the cable. Widening is different between PMT and generator pulses so PMT is used for all channels. The losses in each cable are also monitored using the ratio of the areas of the reflected pulse over the original, taken between the same fractions of each pulse as the timings. This also indicates that any further calibrations of the SD and VCD should be done via their cables to include any effects implicitly.

## Bibliography

[1] RU Beisembaev, EA Beisembaeva, OD Dalkarov, VA Ryabov, AV Stepanov, NG Vildanov, MI Vildanova, VV Zhukov, KA Baigarin, D Beznosko, TX Sadykov, NS Suleymenov, "The 'Horizon-T' Experiment: Extensive Air Showers Detection," *arXiv:1605.05179 [physics.ins-det],* May 17 2016.

[2] D. Beznosko et al., "Horizon-T Extensive Air Showers detector system operations and performance," in *PoS(ICHEP2016)784, proceedings of ICHEP2016*, Chicago, 2016.

[3] Duspayev et al., "The distributed particle detectors and data acquisition modules for Extensive Air Shower measurements at "Horizon-T KZ" experiment," in *PoS(PhotoDet2015)056, in proceedings to PhotoDet2015 conference*, Moscow, 2015.

[4] L. J. Bignell at al., "Characterization and Modeling of a Water-based Liquid Scintillator," *Journal of Instrumentation, IOP Publishing,* vol. 10, p. 12009, 12/2015.

[5] D. Beznosko, "Performance of Water-based Liquid Scintillator," *American Physical Society, APS,* April Meeting 2013, April 13-16, 2013.

[6] Adil Baitenov, Alexander Iakovlev, Dmitriy Beznosko, "Technical manual: a survey of scintillating medium for high-energy particle detection," *arXiv:1601.00086,* 2016/1/1.

[7] MELZ-FEU, 4922-y pr-d, 4c5, Zelenograd, g. Moskva, Russia, 124482 (http://www.melz-feu.ru).

[8] Hamamatsu Corporation, 360 Foothill Road, PO Box 6910, Bridgewater, NJ 08807-0919, USA; 314-5,Shimokanzo, Toyooka-village, Iwatagun,Shizuoka-ken, 438-0193 Japan.

[9] SpetsKabel Inc., 6/1-5 Birusinka St., Moscow, Russia. http://www.spetskabel.ru/.

[10] D. Beznosko, "Novel multi-pixel silicon photon detectors and applications in T2K," *arXiv:0910.4429,* 2009.

[11] T Beremkulov at al., "Random Number Hardware Generator Using Geiger-Mode Avalanche Photo Detector," *arXiv:1501.05521,* Jan 2015.

[12] S Assylbekov et al., "The T2K ND280 off-axis pi–zero detector," *Nuclear Instruments and Methods in Physics Research Section A,* vol. 686, pp. 48-63, 2012/9/11.

[13] CAEN S.p.A. Via della Vetraia, 11, 55049 Viareggio Lucca, Italy. http://caen.it..

[14] "numpy.trapz," [Online]. Available: http://docs.scipy.org/doc/numpy-1.10.1/reference/generated/numpy.trapz.html.

[15] R. Brun, F. Rademakers, "ROOT: An object oriented data analysis framework," *Nucl. Instrum. Meth. A,* vol. 389, p. 81–86, 1997.

[16] D. Beznosko, A. Dyshkant, C.K. Jung, C. McGrew, A. Pla-Dalmau, V. Rykalin, "MRS Photodiode Coupling with Extruded Scintillator via Y7 and Y11 WLS Fibers," February FERMILAB-FN-0796, 2007.

[17] S Assylbekov et al., "The T2K ND280 off-axis pi–zero detector," *Nuclear Instruments and Methods in Physics Research Section A,* vol. 686, pp. 48-63, 2012/9/11.

[18] D. Beznosko, G. Blazey, A. Dyshkant, V. Rykalin, J. Schellpffer and V. Zutshi, "Modular Design for Narrow Scintillating Cells with MRS Photodiodes in Strong Magnetic Field for ILC Detector," Nuclear Instruments and Methods in Physics Research Section A, Volume 564, pages 178-184, 2006/8/1.

[19] K. Abe at al, ""The T2K Experiment"," NIM A Jun 06, 2011 NIMA-D-11-00462R1, doi: 10.1016/j.nima.2011.06.067, arXiv:1106.1238 [physics.ins-det].

[20] D. Beznosko, G. Blazey, A. Dyshkant, V. Rykalin, V. Zutshi, "Effects of the Strong Magnetic Field on LED, Extruded Scintillator and MRS Photodiode," *NIM A,* vol. 553, pp. 438-447, 2005.

[21] D. Beznosko, T. Beremkulov, A. Duspayev, A. Iakovlev, "Random Number Hardware Generator Using Geiger-Mode Avalanche Photo Detector," in *PoS(PhotoDet2015)049, in proceedings to PhotoDet2015*, Moscow, 2015.

[22] M. Yessenov, A. Duspayev, T. Beremkulov, D. Beznosko, A. Iakovlev, M.I. Vildanova, K. Yelshibekov, V.V. Zhukov, "Glass-based charged particle detector performance for Horizon-T EAS detector system," *arxiv:1703.07919,* 03/2017.

[23] A. Batyrkhanov , D. Beznosko, A. Iakovlev. "Optimization of the Liquid Scintillator Composition for Radiation Monitoring Detectors", Science Today - Materials Today: Volume 5, Issue 11, Part 1, 2018, Pages 22770-22775. https://doi.org/10.1016/j.matpr.2018.07.089

[24] Rashid Beisembaev, Dmitriy Beznosko, Kanat Baigarin, Ayan Batyrkhanov, Elena Beisembaeva, Oleg Dalkarov, Alexander Iakovlev, Vladimi Ryabov, Turlan Sadykov, Sergei Shaulov, Marina Vildanova, Tileubek Uakhitov, Valeriy Zhukov, "Extensive Air Showers with Unusual Spatial and Temporal Structure", EPJ Web of Conferences, Volume 208, 06002 (2019) https://doi.org/10.1051/epjconf/201920806002 ISVHECRI 2018

[25] Rashid Beisembaev, Dmitriy Beznosko, Kanat Baigarin, Ayan Batyrkhanov, Elena Beisembaeva, Oleg Dalkarov, Alexander Iakovlev, Vladimi Ryabov, Turlan Sadykov, Sergei Shaulov, Marina Vildanova, Tileubek Uakhitov, Valeriy Zhukov, "Performance of Horizon-10T detector system in Physics Run 1", EPJ Web of Conferences, Volume 208, 08008 (2019) https://doi.org/10.1051/epjconf/201920808008 ISVHECRI 2018

[26] Tileubek Uakhitov, Ayan Batyrkhanov, Dmitriy Beznosko, Alexander Iakovlev, Shotan Jakupov, Amir Kaipiyev, Arailym Konysbayeva, Alikhan Yeltokov, Kamilya Yessimbet and Valeriy Zhukov, "HT-KZ detector system: R&D considerations and prototype performance", EPJ Web of Conferences, Volume 208, 08009 (2019) https://doi.org/10.1051/epjconf/201920808009 ISVHECRI 2018

[27] R.Beisembaev, D.Beznosko, A.Batyrkhanov, K.Baigarin, E.Beisembaeva, O.Dalkarov, A.Iakovlev, V.Ryabov, T.Sadykov, S.Shaulov, T.Uakhitov, M.Vildanova, A.Yeltokov, V.Zhukov, The unusual structure detection in Extensive air shower events at Horizon-8T cosmic rays detector system, accepted to PoS (ICHEP2018)208.